# A Incidência Final dos Tributos Indiretos no Brasil: Estimativa Usando a Matriz de Insumo-Produto 2015

*Rozane Bezerra de Siqueira\*, José Ricardo Bezerra Nogueira\*, Carlos Feitosa Luna\*\**[1]

(março 2021)

**Resumo**

Os tributos que incidem sobre mercadorias e transações respondem por cerca de 45% da receita tributária total no Brasil Essa arrecadação resulta de um sistema altamente complexo, formado por diversos tributos, com diferentes bases de incidência, e uma multiplicidade de alíquotas. Além disso, em torno de 43% dos tributos sobre produtos incidem sobre o 'consumo intermediário das atividades produtivas', ou seja, sobre insumos. Nesse contexto, as alíquotas tributárias efetivas sobre os bens e serviços produzidos pelos diferentes setores de atividade podem divergir substancialmente das alíquotas estatutárias. Neste estudo, estimamos a incidência final dos tributos indiretos no Brasil usando a matriz brasileira de insumo-produto de 2015 e um método que incorpora os efeitos multissetoriais da tributação de insumos.

**Palavras-chave**: Impostos Indiretos – Tributação de Insumos - Incidência

**Classificação JEL**: H2, H22

**Abstract**

Taxes on goods and services account for about 45% of total tax revenue in Brazil. This tax collection results in a highly complex system, with several taxes, different tax bases, and a multiplicity of rates. Moreover, about 43% of taxes on goods fall on inputs. In this context, the effective tax rates can substantially differ from the legal rates. In this study we estimate the final incidence of indirect taxes in Brazil using the 2015 Brazilian input-output matrix and a method that incorporates the multisector effects of the taxation of inputs.

**Key words**: Indirect Taxes – Taxation of Inputs - Incidence

**JEL Classification**: H2, H22

[1] \* Departamento de Economia, Centro de Ciências Sociais Aplicadas, Universidade Federal de Pernambuco; \*\* Fundação Oswaldo Cruz/FIOCRUZ. Endereço para correspondência: rozane_siqueira@yahoo.com.br; jrbnogueira@yahoo.com.br.

## 1. Introdução

Os tributos que incidem sobre mercadorias e transações – denominados 'tributos indiretos' – respondem por cerca de 45% da receita tributária total no Brasil, equivalente a 14,8% do PIB em 2019 (STN, 2020). Essa arrecadação resulta de um sistema altamente complexo, formado por diversos tributos, com diferentes bases de incidência, e uma multiplicidade de alíquotas. Além disso, de acordo com a *Matriz de Insumo-Produto do Brasil 2015* (IBGE, 2018), em torno de 43% dos tributos sobre produtos incidem sobre o 'consumo intermediário das atividades produtivas', ou seja, sobre insumos. Nesse contexto, a incidência final total dos tributos indiretos fica longe de ser transparente. Em particular, as alíquotas tributárias efetivas sobre os bens e serviços produzidos pelos diferentes setores de atividade podem divergir substancialmente das alíquotas estatutárias.

O objetivo deste estudo é estimar a incidência final dos tributos indiretos no Brasil usando a matriz brasileira de insumo-produto de 2015 e um método que incorpora os efeitos multissetoriais da tributação de insumos. Tal método, proposto em Scutella (1999), permite calcular a incidência em reais, bem com as respectivas alíquotas efetivas, por produto, para cada componente da demanda final: consumo das famílias, consumo do governo, exportações e investimento. Do nosso conhecimento, apenas Siqueira, Nogueira e Souza (2001) estima a incidência efetiva da tributação indireta no Brasil desagregando por categoria de demanda final.[2] Sendo baseado na matriz de insumo-produto de 1995, esse estudo anterior encontra-se bastante defasado. Por outro lado, o debate sobre reforma tributária no Brasil continua bem atual, de forma que novas estimativas podem ser úteis.

Além desta introdução, este artigo é composto de mais três seções. A seção 2 descreve o método e os dados usados para calcular a incidência final dos tributos indiretos no Brasil. A incidência final em reais e as respectivas alíquotas efetivas, por produto e por componente da demanda final, são apresentadas na Seção 3. A seção 4 apresenta comentários finais.

## 2. Método e Dados

Para estimar a incidência final dos tributos indiretos no Brasil, este trabalho utiliza o modelo básico e procedimentos descritos detalhadamente em Scutella (1999). A abordagem é apresentada, de forma resumida, nesta seção, com o objetivo de explicitar as principais hipóteses adotadas, bem como chamar a atenção para algumas escolhas metodológicas que diferem daquelas de Siqueira, Nogueira e Souza (2001).

O método de Scutella (1999) parte das receitas arrecadadas de cada atividade produtiva por meio dos tributos indiretos. Essa incidência inicial ('quem deposita na conta do governo') é denominada *incidência estatutária*. Supõe-se então que as atividades produtivas transferem totalmente os tributos pagos por elas para seus consumidores finais e para as atividades que utilizam seus produtos como insumos.[3] A transferência inicial para outras atividades produtivas é aqui denominada de *incidência de primeiro estágio sobre a demanda intermediária*. Em um segundo estágio, o tributo transferido para uma atividade intermediária é, novamente, repassado para consumidores finais e para outras atividades produtivas, e assim sucessivamente, até que, após muitos estágios, todo tributo recaia sobre a demanda final.

---

[2] Siqueira, Nogueira e Souza (2010) apresenta estimativas da incidência efetiva dos tributos indiretos sobre o consumo das famílias apenas.

[3] Uma discussão sobre as implicações e validade dessa hipótese é apresentada em Siqueira, Nogueira e Souza (2001).

Informação sobre a incidência estatutária dos tributos indiretos, bem como sobre a incidência de primeiro estágio, pode ser obtida diretamente da matriz de insumo-produto brasileira, produzida pelo Instituto Brasileiro de Estatística (IBGE). O método proposto por Scutella (1999) para calcular a *incidência de estágios subsequentes* é descrito a seguir.

Seja **IDI** um vetor que expressa a incidência de primeiro estágio (estágio 1) sobre a demanda intermediária, por setor de atividade, e **A** uma matriz de coeficientes que expressa a oferta intermediária da atividade *i* para a atividade demandante *j* como percentagem da oferta total da atividade *i*.

O montante de tributo que permanece sobre a demanda intermediária no estágio 2 é **IDI**' x **I**, no estágio 3 é **IDI**' x **A**, no estágio 4 é **IDI**' x $\mathbf{A}^2$, no estágio n é **IDI**' x $\mathbf{A}^{n-2}$, onde IDI' é o transposto do vetor da incidência de primeiro estágio sobre a demanda intermediária e I denota a matriz identidade.

Depois de n+2 estágios, o montante de imposto transferido para a demanda final, denominado aqui *incidência de estágios subsequentes*, **IES**, é dado por:

$$\mathbf{IES} = (\mathbf{IDI}' \times \mathbf{I})' \# \mathbf{Z} + (\mathbf{IDI}' \times \mathbf{A})' \# \mathbf{Z} + (\mathbf{IDI}' \times \mathbf{A}^2)' \# \mathbf{Z} + ... + (\mathbf{IDI}' \times \mathbf{A}^n)' \# \mathbf{Z}$$

onde o símbolo # expressa multiplicação elemento por elemento (e não multiplicação matricial) e Z é um vetor que fornece a parcela do imposto sobre a atividade *i* que é transferida para a demanda final em cada estágio (dada pela razão entre a demanda final e a oferta total da atividade *i*).

A expressão acima pode ser reescrita como:

$$\mathbf{IES} = [\mathbf{IDI}' \times (\mathbf{I} + \mathbf{A} + \mathbf{A}^2 + ... + \mathbf{A}^n)]' \# \mathbf{Z}$$

Quando n é infinito, a expressão entre parênteses é uma série geométrica infinita, que pode ser simplificada para:

$$\mathbf{IES} = [\mathbf{IDI}' \times (\mathbf{I} - \mathbf{A})^{-1}]' \# \mathbf{Z}$$

A incidência final total dos impostos indiretos é a soma da incidência de primeiro estágio sobre a demanda final e a incidência de estágios subsequentes.

Ressalte-se, no entanto, que a aplicação direta da expressão acima não garante que um imposto sobre 'serviços margens', ou sobre insumos usados por atividades margens, seja totalmente incorporado no preço final dos produtos das atividades que usam serviços margem. Scutella (1999) explica que isso ocorre devido a forma como os fluxos envolvendo atividades margens são registrados nas matrizes de insumo-produto.

Atividades margens são serviços que adicionam valor a um produto em seu caminho de uma atividade para outra na cadeia de produção/comercialização. Na matriz de insumo-produto brasileira as atividades explicitamente identificadas como ofertantes de serviços margens são: transporte terrestre de carga, transporte aquaviário e comércio por atacado e varejo.[4] Para

---

[4] Deve-se observar que apenas uma parcela da oferta total dessas atividades corresponde a serviços margens. Na matriz de insumo-produto de 2015, margem de comércio representa 89,5% da oferta total da

garantir que a tributação de serviços margens se traduza totalmente em aumento de preço de bens finais, Scutella (1999) redistribui a oferta de serviços margens e os tributos que incidem sobre esses serviços entre as atividades produtivas que utilizam serviços margens. O mesmo procedimento é adotado neste estudo.[5]

O método básico descrito acima para calcular a incidência final dos tributos indiretos pode ser aplicado em um modelo de curto prazo ou em um modelo de longo prazo. Os dois modelos se diferenciam quanto ao tratamento do investimento. No modelo de longo prazo, o investimento é considerado um insumo intermediário usado pelas atividades produtivas e a coluna 'formação bruta de capita fixo' da matriz de insumo-produto é deslocada para demanda intermediária, enquanto no modelo de curto prazo a formação bruta de capital fixo permanece na demanda final. Diferentemente de Siqueira, Nogueira e Souza (2001), o presente estudo adota a abordagem de curto prazo, buscando explicitar a incidência de tributos sobre o investimento.

A matriz de insumo produto usada neste trabalho é a última disponível para o Brasil, referente ao ano de 2015 (IBGE, 2018). Como já mencionado, a matriz contém informação sobre a incidência estatutária dos tributos indiretos, bem como sobre incidência de primeiro estágio desses tributos.[6] Essa informação, apresentada para 67 atividades e 127 produtos, é para o agregado de tributos indiretos líquidos de subsídios. De acordo o relatório metodológico do Sistema de Contas Nacionais – referência 2010 (IBGE, 2016), os tributos sobre cada produto são estimados pelo IBGE considerando as alíquotas e as bases de incidência de cada tipo de tributo conforme a legislação em vigor, bem como levando em conta estimativas de sonegação e contrabando. Ao final do processo, os valores estimados de cada tributo, por produto, são ajustados ao valor arrecadado pelo governo.[7]

A tabela 1 relaciona os principais tributos indiretos incorporados na matriz de insumo-produto de 2015, apresentando também a receita arrecadada naquele ano. O ICMS é o mais importante em termos de arrecadação, respondendo por 47,7% da receita proveniente da tributação que incide sobre mercadorias e transações. Em seguida, vem a Cofins, contribuindo com 24% da receita.[8]

---

atividade 'comércio por atacado e varejo'. Por sua vez, margem de transporte responde por 32,2% e 7,9%, respectivamente, das atividades 'transporte terrestre de carga' e 'transporte aquaviário'.

[5] Ver Scutella (1999) para uma descrição passo a passo desse procedimento.

[6] A informação sobre a incidência de primeiro estágio na matriz de insumo-produto consiste das tabelas intituladas "destino dos impostos sobre produtos nacionais" e "destino dos impostos sobre produtos importados".

[7] Ver o Apêndice 3 do referido relatório.

[8] Ressalte-se que o ITBI passou a ser tratado como imposto sobre produto pelo IBGE a partir da série Sistema de Contas Nacionais – 2010. Anteriormente, esse tributo era tratado como imposto sobre capital. Na matriz de insumo-produto, o ITBI é alocado integralmente aos produtos da construção civil (IBGE, 2016).

**Tabela 1: Tributos sobre Mercadorias e Transações, 2015**

| Tributo | R$(milhões) | % |
|---|---|---|
| Imposto sobre Circulação de Mercadorias e Serviços – ICMS | 396.513 | 47,7 |
| Contribuição para o Financiamento da Seguridade Social – Cofins | 199.876 | 24,0 |
| Imposto sobre Serviços de Qualquer Natureza – ISS | 58.083 | 7,0 |
| Contribuição para o PIS | 39.825 | 4,8 |
| Imposto sobre Produtos Industrializados – IPI | 48.048 | 5,8 |
| Impostos sobre o comércio exterior | 39.969 | 4,8 |
| Imposto sobre Operações Financeiras – IOF | 34.681 | 4,2 |
| Imposto de Transmissão de Bens Imóveis – ITBI | 11.106 | 1,3 |
| Cide-Combustíveis | 3.271 | 0,4 |
| **Total** | **831.374** | **100,0** |

Fonte: "Carga Tributária no Brasil: 2015", Receita Federal, 2016.

## 3. Resultados

A tabela 2 apresenta a incidência estatutária e a incidência de primeiro estágio dos tributos indiretos (líquidos de subsídios) no Brasil em 2015. Essa tabela é elaborada a partir das tabelas "Destino dos impostos sobre produtos nacionais" e "Destino dos impostos sobre produtos importados" da matriz de insumo-produto de 2015 (IBGE, 2018). Note-se, porém, que para alguns produtos a incidência apresentada na tabela 2 é levemente diferente daquela obtida diretamente das tabelas da Matriz, devido à redistribuição dos tributos que incidem sobre serviços margens para as atividades usuárias desses serviços, conforme explicado na seção 2.[9]

Como já mencionado na introdução, 43,3% dos tributos sobre mercadorias e transações tem como destino inicial a demanda intermediária. Do montante que incide sobre a demanda final, 88,6% recaem sobre o consumo das famílias e 11% incidem sobre a formação bruta de capital fixo. Da incidência total sobre a formação bruta de capital fixo, 52,9% são sobre produtos do setor de construção.

A tabela 3 apresenta a incidência final dos tributos indiretos no Brasil em 2015. Do total arrecadado, 76% tem incidência final sobre o consumo das famílias e 13,9% sobre a formação bruta de capital fixo. Apesar das exportações serem legalmente isentas (incluindo seus insumos diretos), como refletido na incidência de primeiro estágio (tabela 2), as estimativas apresentadas na tabela 3 indicam que 6,7% do total de tributos indiretos acaba recaindo sobre esse componente da demanda final.[10]

A partir da última coluna da tabela 3 pode-se verificar que em torno de 11% da tributação indireta incide sobre Produtos Alimentares. Destaca-se ainda a incidência sobre os produtos das seguintes atividades: Construção (7,2% do total de tributos indiretos), Refino do Petróleo (7,1%), Intermediação Financeira e Seguros (6,6%), Indústria Automobilística (5,5%), Telecomunicações (4,5%), e Energia Elétrica, Gás Natural e outras Utilidades (4,4%).

---

[9] Nas tabelas de resultado aqui apresentadas optamos por excluir as colunas 'Consumo das Instituições Sem Fim Lucrativo a Serviço das Famílias - ISFLSF' e 'Variação de estoque', que são dois componentes da demanda final.

[10] A diferença entre a soma das quatro primeiras colunas da tabela 3 e a coluna 'total da demanda final' representa a incidência sobre os componentes 'consumo das ISFLSF' e 'variação de estoque'.

A tabela 4 mostra as alíquotas efetivas associadas à tributação indireta no Brasil em 2015. Essas alíquotas são calculadas com o tributo 'por fora', ou seja, representam a divisão de cada montante apresentado na tabela 3 pela respectiva despesa deduzida do valor do tributo. Talvez o aspecto que chama mais a atenção nessa tabela seja a enorme diferenciação das alíquotas entre os produtos das diferentes atividades produtivas.

A alíquota média sobre o consumo das famílias é 19,1%, porém, para os produtos de 18 atividades, a alíquota efetiva é superior a 30%, com destaque para Energia Elétrica e Gás Natural (48,7%), produtos do Refino do Petróleo' (45,0%) e Telecomunicações (42,2%). Em contraste, as alíquotas sobre serviços consumidos pelas famílias são em geral bem inferiores à alíquota efetiva média dos tributos indiretos. Por exemplo, sobre serviços prestados por profissionais (jurídicos, contabilidade, publicidade, administrativos), a alíquota efetiva é, em média, cerca de 12%. Já as alíquotas efetivas sobre os serviços de educação e saúde privados são estimadas em 4,0% e 5,1%, respectivamente.[11]

Olhando para os outros componentes da demanda final, pode-se verificar que a alíquota efetiva média sobre a formação bruta de capital fixo é 11,4%, em grande medida refletindo a alíquota efetiva sobre os produtos da atividade de Construção (11,2%). Por sua vez, a alíquota efetiva média sobre exportações é 7,3%, que é essencialmente uma consequência da tributação de insumos indiretos, e não um resultado intencional da política do governo, como comentado acima.

---

[11] Cabe lembrar que os tributos indiretos estimados pelo IBGE na construção da matriz de insumo-produto são 'líquidos de subsídios'. O relatório metodológico Sistema de Contas Nacionais – referência 2010 (IBGE, 2016) esclarece que "os subsídios à produtos consistem das despesas de equalizações ou subvenções econômicas destinadas à redução dos preços ou aumento da oferta de produtos específicos". É sabido que instituições que prestam serviços de saúde e educação são particularmente beneficiadas por subvenções.

**Tabela 2 - Incidência Estatutária e Incidência de Primeiro Estágio, 2015 (R$ milhões)**

| Atividade | Incidência Estatutária | Incidência de Primeiro Estágio | | | | |
|---|---|---|---|---|---|---|
| | | Demanda Intermediária | Demanda Final | | | |
| | | | Exportação de bens e serviços | Consumo do governo | Consumo das famílias | Formação bruta de capital fixo |
| Agricultura, inclusive o apoio à agricultura e a pós-colheita | 5.663 | 2.540 | 0 | 6 | 3.115 | 2 |
| Pecuária, inclusive o apoio à pecuária | 3.984 | 2.604 | 0 | 1 | 948 | 431 |
| Produção florestal; pesca e aquicultura | 3.357 | 1.233 | 0 | 2 | 2.093 | 29 |
| Extração de carvão mineral e de minerais não-metálicos | 1.793 | 1.793 | 0 | 0 | 0 | 0 |
| Extração de petróleo e gás, inclusive as atividades de apoio | 2.693 | 2.538 | 0 | 0 | 0 | 155 |
| Extração de minério de ferro, inclusive beneficiamentos e a aglomeração | 221 | 221 | 0 | 0 | 0 | 0 |
| Extração de minerais metálicos não-ferrosos, inclusive beneficiamentos | 531 | 531 | 0 | 0 | 0 | 0 |
| Abate e produtos de carne, inclusive os produtos do laticínio e da pesca | 29.969 | 2.280 | 0 | 12 | 27.678 | 0 |
| Fabricação e refino de açúcar | 1.344 | 139 | 0 | 0 | 1.205 | 0 |
| Outros produtos alimentares | 37.752 | 5.055 | 0 | 16 | 32.681 | 0 |
| Fabricação de bebidas | 25.124 | 8.036 | 0 | 0 | 17.088 | 0 |
| Fabricação de produtos do fumo | 7.932 | 95 | 12 | 0 | 7.825 | 0 |
| Fabricação de produtos têxteis | 10.153 | 4.782 | 0 | 0 | 5.371 | 0 |
| Confecção de artefatos do vestuário e acessórios | 22.787 | 1.142 | 0 | 2 | 21.643 | 0 |
| Fabricação de calçados e de artefatos de couro | 10.373 | 397 | 21 | 0 | 9.955 | 0 |
| Fabricação de produtos da madeira | 2.904 | 2.552 | 0 | 0 | 348 | 4 |
| Fabricação de celulose, papel e produtos de papel | 8.601 | 5.579 | 0 | 0 | 3.022 | 0 |
| Impressão e reprodução de gravações | 2.944 | 2.888 | 0 | 0 | 56 | 0 |
| Refino de petróleo e coquerias | 87.273 | 51.990 | 0 | 0 | 35.284 | 0 |
| Fabricação de biocombustíveis | 12.834 | 2.691 | 0 | 0 | 10.143 | 0 |
| Fabricação de químicos orgânicos e inorgânicos, resinas e elastômeros | 15.671 | 15.666 | 0 | 0 | 5 | 0 |
| Fabricação de defensivos, desinfestantes, tintas e químicos diversos | 9.889 | 9.634 | 0 | 0 | 255 | 0 |
| Fabricação de produtos de limpeza, cosméticos e higiene pessoal | 20.509 | 2.305 | 0 | 0 | 18.204 | 0 |
| Fabricação de produtos farmoquímicos e farmacêuticos | 18.225 | 5.407 | 0 | 658 | 12.159 | 0 |
| Fabricação de produtos de borracha e de material plástico | 12.874 | 8.881 | 0 | 0 | 3.994 | 0 |
| Fabricação de produtos de minerais não-metálicos | 12.124 | 11.261 | 0 | 0 | 863 | 0 |
| Produção de ferro-gusa/ferroligas, siderurgia e tubos de aço | 5.620 | 5.577 | 0 | 0 | 43 | 0 |
| Metalurgia de metais não-ferosos e a fundição de metais | 3.277 | 3.257 | 0 | 0 | 20 | 0 |
| Fabricação de produtos de metal, exceto máquinas e equipamentos | 12.636 | 9.057 | 67 | 0 | 3.080 | 432 |
| Fabricação de equipamentos de informática, eletrônicos e ópticos | 27.725 | 8.285 | 0 | 0 | 15.751 | 3.689 |
| Fabricação de máquinas e equipamentos elétricos | 18.172 | 7.374 | 0 | 0 | 9.048 | 1.750 |
| Fabricação de máquinas e equipamentos mecânicos | 15.730 | 6.609 | 0 | 0 | 1.896 | 7.225 |
| Fabricação de automóveis, caminhões e ônibus, exceto peças | 33.976 | 650 | 0 | 0 | 27.994 | 5.332 |
| Fabricação de peças e acessórios para veículos automotores | 8.601 | 8.601 | 0 | 0 | 0 | 0 |
| Fabricação de outros equipamentos de transporte, exceto veículos automotores | 6.276 | 960 | 0 | 0 | 4.299 | 1.017 |
| Fabricação de móveis e de produtos de indústrias diversas | 22.043 | 5.573 | 0 | 0 | 14.362 | 2.108 |
| Manutenção, reparação e instalação de máquinas e equipamentos | 4.287 | 3.698 | 0 | 0 | 32 | 557 |
| Energia elétrica, gás natural e outras utilidades | 46.751 | 22.539 | 0 | 0 | 24.212 | 0 |
| Água, esgoto e gestão de resíduos | 3.159 | 2.002 | 0 | 0 | 1.156 | 0 |
| Construção | 33.436 | 5.813 | 0 | 0 | 0 | 27.623 |
| Comércio por atacado e varejo | 321 | 196 | 0 | 0 | 125 | 0 |
| Transporte terrestre | 13.471 | 6.918 | 0 | 0 | 6.553 | 0 |
| Transporte aquaviário | 3.262 | 3.152 | 0 | 0 | 110 | 0 |
| Transporte aéreo | 1.831 | 1.342 | 0 | 0 | 489 | 0 |
| Armazenamento, atividades auxiliares dos transportes e correio | 7.322 | 5.637 | 0 | 0 | 1.685 | 0 |
| Alojamento | 3.237 | 2.331 | 0 | 0 | 905 | 0 |
| Alimentação | 24.541 | 3.105 | 0 | 0 | 21.436 | 0 |
| Edição e edição integrada à impressão | 800 | 353 | 0 | 0 | 447 | 0 |
| Atividades de televisão, rádio, cinema e gravação de som e imagem | 1.832 | 1.708 | 0 | 0 | 124 | 0 |
| Telecomunicações | 37.688 | 9.497 | 0 | 0 | 28.191 | 0 |
| Desenvolvimento de sistemas e outros serviços de informação | 5.858 | 4.147 | 0 | 0 | 28 | 1.683 |
| Intermediação financeira, seguros e previdência complementar | 70.143 | 39.094 | 0 | 0 | 31.049 | 0 |
| Atividades imobiliárias | 1.144 | 844 | 0 | 0 | 300 | 0 |
| Atividades jurídicas, contábeis, consultoria e sedes de empresas | 10.327 | 9.542 | 0 | 0 | 785 | 0 |
| Serviços de arquitetura, engenharia, testes/análises técnicas e P & D | 3.921 | 3.727 | 0 | 0 | 39 | 155 |
| Outras atividades profissionais, científicas e técnicas | 6.242 | 6.204 | 0 | 0 | 38 | 0 |
| Aluguéis não-imobiliários e gestão de ativos de propriedade intelectual | 11.095 | 10.935 | 0 | 0 | 160 | 0 |
| Outras atividades administrativas e serviços complementares | 9.868 | 9.313 | 0 | 0 | 555 | 0 |
| Atividades de vigilância, segurança e investigação | 1.784 | 1.766 | 0 | 0 | 18 | 0 |
| Administração pública, defesa e seguridade social | 0 | 0 | 0 | 0 | 0 | 0 |
| Educação pública | 0 | 0 | 0 | 0 | 0 | 0 |
| Educação privada | 2.036 | 207 | 0 | 0 | 1.829 | 0 |
| Saúde pública | 0 | 0 | 0 | 0 | 0 | 0 |
| Saúde privada | 6.092 | 442 | 0 | 1.206 | 4.444 | 0 |
| Atividades artísticas, criativas e de espetáculos | 6.397 | 87 | 0 | 0 | 6.310 | 0 |
| Organizações associativas e outros serviços pessoais | 1.761 | 954 | 0 | 0 | 807 | 0 |
| Serviços domésticos | 0 | 0 | 0 | 0 | 0 | 0 |
| **Total** | **840.186** | **363.735** | **100** | **1.903** | **422.257** | **52.191** |

**Tabela 3: Incidência Final dos Tributos Indiretos no Brasil, 2015 (R$milhões)**

| Atividades | Exportação de Bens e Serviços | Consumo do Governo | Consumo das Famílias | Formação Bruta de Capital Fixo | Total da Demanda Final |
|---|---|---|---|---|---|
| Agricultura, inclusive o apoio à agricultura e a pós-colheita | 6.285 | 10 | 7.883 | 24 | 13.970 |
| Pecuária, inclusive o apoio à pecuária | 119 | 1 | 2.243 | 1.219 | 3.636 |
| Produção florestal; pesca e aquicultura | 98 | 2 | 2.988 | 63 | 3.123 |
| Extração de carvão mineral e de minerais não-metálicos | 155 | 0 | 0 | 0 | 210 |
| Extração de petróleo e gás, inclusive as atividades de apoio | 2.060 | 0 | 0 | 609 | 2.880 |
| Extração de minério de ferro, inclusive beneficiamentos e a aglomeração | 2.192 | 0 | 0 | 0 | 2.257 |
| Extração de minerais metálicos não-ferrosos, inclusive beneficiamentos | 592 | 0 | 0 | 0 | 689 |
| Abate e produtos de carne, inclusive os produtos do laticínio e da pesca | 2.730 | 20 | 38.739 | 0 | 41.501 |
| Fabricação e refino de açúcar | 1.659 | 0 | 1.739 | 0 | 3.254 |
| Outros produtos alimentares | 2.187 | 26 | 45.319 | 0 | 47.498 |
| Fabricação de bebidas | 426 | 0 | 24.211 | 0 | 24.385 |
| Fabricação de produtos do fumo | 247 | 0 | 8.305 | 0 | 8.569 |
| Fabricação de produtos têxteis | 254 | 0 | 8.353 | 0 | 8.457 |
| Confecção de artefatos do vestuário e acessórios | 66 | 3 | 25.516 | 0 | 25.517 |
| Fabricação de calçados e de artefatos de couro | 458 | 0 | 11.731 | 0 | 12.149 |
| Fabricação de produtos da madeira | 920 | 0 | 631 | 17 | 1.352 |
| Fabricação de celulose, papel e produtos de papel | 2.900 | 0 | 4.635 | 0 | 7.366 |
| Impressão e reprodução de gravações | 10 | 0 | 137 | 0 | 106 |
| Refino de petróleo e coquerias | 2.675 | 0 | 57.195 | 0 | 59.385 |
| Fabricação de biocombustíveis | 222 | 0 | 12.595 | 0 | 12.554 |
| Fabricação de químicos orgânicos e inorgânicos, resinas e elastômeros | 2.317 | 0 | 8 | 0 | 2.767 |
| Fabricação de defensivos, desinfestantes, tintas e químicos diversos | 992 | 0 | 443 | 0 | 1.568 |
| Fabricação de produtos de limpeza, cosméticos/perfumaria e higiene pessoal | 264 | 0 | 22.436 | 0 | 22.695 |
| Fabricação de produtos farmoquímicos e farmacêuticos | 289 | 1.320 | 17.193 | 0 | 18.793 |
| Fabricação de produtos de borracha e de material plástico | 1.110 | 0 | 5.815 | 0 | 6.638 |
| Fabricação de produtos de minerais não-metálicos | 1.177 | 0 | 1.337 | 0 | 2.492 |
| Produção de ferro-gusa/ferroligas, siderurgia e tubos de aço sem costura | 4.001 | 0 | 64 | 10 | 3.852 |
| Metalurgia de metais não-ferosos e a fundição de metais | 2.396 | 0 | 67 | 0 | 2.492 |
| Fabricação de produtos de metal, exceto máquinas e equipamentos | 830 | 0 | 4.862 | 1.607 | 7.251 |
| Fabricação de equipamentos de informática, produtos eletrônicos e ópticos | 302 | 0 | 20.130 | 8.074 | 28.283 |
| Fabricação de máquinas e equipamentos elétricos | 929 | 0 | 12.429 | 4.181 | 17.588 |
| Fabricação de máquinas e equipamentos mecânicos | 1.544 | 0 | 2.249 | 14.538 | 18.213 |
| Fabricação de automóveis, caminhões e ônibus, exceto peças | 1.567 | 0 | 35.159 | 10.088 | 46.444 |
| Fabricação de peças e acessórios para veículos automotores | 1.891 | 0 | 0 | 0 | 1.874 |
| Fabricação de outros equipamentos de transporte, exceto veículos automotores | 1.316 | 0 | 4.898 | 1.894 | 8.155 |
| Fabricação de móveis e de produtos de indústrias diversas | 312 | 0 | 19.990 | 3.442 | 23.684 |
| Manutenção, reparação e instalação de máquinas e equipamentos | 190 | 0 | 70 | 1.475 | 1.735 |
| Energia elétrica, gás natural e outras utilidades | 0 | 0 | 37.193 | 0 | 37.193 |
| Água, esgoto e gestão de resíduos | 0 | 0 | 2.713 | 0 | 2.741 |
| Construção | 284 | 0 | 0 | 59.917 | 60.200 |
| Comércio por atacado e varejo | 535 | 109 | 18.174 | 1.886 | 20.704 |
| Transporte terrestre | 175 | 5 | 17.763 | 363 | 18.307 |
| Transporte aquaviário | 974 | 0 | 379 | 2 | 1.355 |
| Transporte aéreo | 562 | 0 | 1.421 | 0 | 1.983 |
| Armazenamento, atividades auxiliares dos transportes e correio | 672 | 0 | 3.564 | 0 | 4.236 |
| Alojamento | 478 | 0 | 1.661 | 0 | 2.139 |
| Alimentação | 254 | 0 | 31.749 | 0 | 32.002 |
| Edição e edição integrada à impressão | 43 | 0 | 1.196 | 0 | 1.244 |
| Atividades de televisão, rádio, cinema e gravação/edição de som e imagem | 34 | 0 | 248 | 0 | 283 |
| Telecomunicações | 137 | 0 | 37.380 | 0 | 37.517 |
| Desenvolvimento de sistemas e outros serviços de informação | 196 | 0 | 48 | 4.932 | 5.178 |
| Intermediação financeira, seguros e previdência complementar | 1.249 | 176 | 54.242 | 0 | 55.666 |
| Atividades imobiliárias | 42 | 0 | 4.009 | 0 | 4.051 |
| Atividades jurídicas, contábeis, consultoria e sedes de empresas | 959 | 0 | 1.715 | 0 | 2.674 |
| Serviços de arquitetura, engenharia, testes/análises técnicas e P & D | 684 | 0 | 64 | 2.398 | 3.147 |
| Outras atividades profissionais, científicas e técnicas | 260 | 0 | 182 | 0 | 441 |
| Aluguéis não-imobiliários e gestão de ativos de propriedade intelectual | 397 | 0 | 610 | 0 | 1.007 |
| Outras atividades administrativas e serviços complementares | 642 | 0 | 1.261 | 0 | 3.681 |
| Atividades de vigilância, segurança e investigação | 0 | 0 | 41 | 0 | 41 |
| Administração pública, defesa e seguridade social | 0 | 15.654 | 0 | 0 | 15.654 |
| Educação pública | 0 | 3.662 | 0 | 0 | 3.662 |
| Educação privada | 2 | 0 | 4.087 | 0 | 4.089 |
| Saúde pública | 0 | 4.046 | 0 | 0 | 4.046 |
| Saúde privada | 5 | 2.040 | 8.285 | 0 | 10.511 |
| Atividades artísticas, criativas e de espetáculos | 45 | 0 | 6.905 | 0 | 7.149 |
| Organizações associativas e outros serviços pessoais | 0 | 0 | 4.107 | 0 | 5.899 |
| Serviços domésticos | 0 | 0 | 0 | 0 | 0 |
| **Total** | **56.314** | **27.076** | **638.367** | **116.739** | **840.186** |

**Tabela 4 - Alíquotas Efetivas da Tributação Indireta sobre a Demanda Final, Brasil 2015 (%)**

| Atividade | Exportação | Consumo do Governo | Consumo das Famílias | Formação Bruta de Capital Fixo | Total da Damanda Final |
|---|---|---|---|---|---|
| Agricultura, inclusive o apoio à agricultura e a pós-colheita | 5,4 | ND | 8,9 | ND | 7,0 |
| Pecuária, inclusive o apoio à pecuária | 5,9 | ND | 10,3 | 9,2 | 9,6 |
| Produção florestal; pesca e aquicultura | 4,3 | ND | 14,2 | ND | 13,3 |
| Extração de carvão mineral e de minerais não-metálicos | 8,1 | ND | ND | ND | 8,1 |
| Extração de petróleo e gás, inclusive as atividades de apoio | 5,3 | ND | ND | 7,1 | 5,6 |
| Extração de minério de ferro, inclusive beneficiamentos e a aglomeração | 4,7 | ND | ND | ND | 4,7 |
| Extração de minerais metálicos não-ferrosos, inclusive beneficiamentos | 6,9 | ND | ND | ND | 6,9 |
| Abate e produtos de carne, inclusive os produtos do laticínio e da pesca | 5,0 | ND | 17,6 | ND | 15,1 |
| Fabricação e refino de açúcar | 6,5 | ND | 21,1 | ND | 10,3 |
| Outros produtos alimentares | 5,3 | ND | 18,9 | ND | 16,9 |
| Fabricação de bebidas | 12,2 | ND | 41,3 | ND | 40,6 |
| Fabricação de produtos do fumo | 3,4 | ND | 55,8 | ND | 37,7 |
| Fabricação de produtos têxteis | 9,9 | ND | 27,7 | ND | 27,1 |
| Confecção de artefatos do vestuário e acessórios | 3,4 | ND | 22,4 | ND | 22,4 |
| Fabricação de calçados e de artefatos de couro | 4,0 | ND | 25,5 | ND | 21,5 |
| Fabricação de produtos da madeira | 12,4 | ND | 27,6 | ND | 16,7 |
| Fabricação de celulose, papel e produtos de papel | 11,2 | ND | 32,1 | ND | 18,9 |
| Impressão e reprodução de gravações | ND | ND | ND | ND | ND |
| Refino de petróleo e coquerias | 17,2 | ND | 45,0 | ND | 42,5 |
| Fabricação de biocombustíveis | 7,1 | ND | 36,4 | ND | 36,9 |
| Fabricação de químicos orgânicos e inorgânicos, resinas e elastômeros | 11,8 | ND | ND | ND | 11,8 |
| Fabricação de defensivos, desinfestantes, tintas e químicos diversos | 14,4 | ND | 33,9 | ND | 17,1 |
| Fabricação de produtos de limpeza, cosméticos/perfumaria e higiene pessoal | 6,6 | ND | 35,1 | ND | 33,4 |
| Fabricação de produtos farmoquímicos e farmacêuticos | 6,5 | 12,9 | 22,1 | ND | 20,4 |
| Fabricação de produtos de borracha e de material plástico | 12,9 | ND | 41,2 | ND | 32,4 |
| Fabricação de produtos de minerais não-metálicos | 15,5 | ND | 43,8 | ND | 23,8 |
| Produção de ferro-gusa/ferroligas, siderurgia e tubos de aço sem costura | 10,8 | ND | ND | ND | 11,0 |
| Metalurgia de metais não-ferosos e a fundição de metais | 9,5 | ND | ND | ND | 9,6 |
| Fabricação de produtos de metal, exceto máquinas e equipamentos | 12,2 | ND | 30,6 | 15,4 | 22,2 |
| Fabricação de equipamentos de informática, produtos eletrônicos e ópticos | 6,9 | ND | 31,9 | 12,8 | 22,2 |
| Fabricação de máquinas e equipamentos elétricos | 10,0 | ND | 36,7 | 17,2 | 25,9 |
| Fabricação de máquinas e equipamentos mecânicos | 6,3 | ND | 39,7 | 12,4 | 12,5 |
| Fabricação de automóveis, caminhões e ônibus, exceto peças | 6,3 | ND | 31,0 | 13,4 | 22,4 |
| Fabricação de peças e acessórios para veículos automotores | 11,9 | ND | ND | ND | 11,9 |
| Fabricação de outros equipamentos de transporte, exceto veículos automotores | 4,4 | ND | 35,8 | 9,4 | 12,6 |
| Fabricação de móveis e de produtos de indústrias diversas | 6,2 | ND | 21,9 | 15,9 | 20,2 |
| Manutenção, reparação e instalação de máquinas e equipamentos | 8,0 | ND | ND | 12,8 | 12,0 |
| Energia elétrica, gás natural e outras utilidades | ND | ND | 48,7 | ND | 48,7 |
| Água, esgoto e gestão de resíduos | ND | ND | 11,0 | ND | 10,9 |
| Construção | 6,0 | ND | ND | 11,2 | 11,2 |
| Comércio por atacado e varejo | 30,7 | ND | 30,9 | 30,7 | 30,8 |
| Transporte terrestre | 12,7 | ND | 20,1 | 12,7 | 19,8 |
| Transporte aquaviário | 16,0 | ND | 22,5 | ND | 17,4 |
| Transporte aéreo | 9,4 | ND | 14,3 | ND | 12,5 |
| Armazenamento, atividades auxiliares dos transportes e correio | 7,4 | ND | 14,0 | ND | 12,3 |
| Alojamento | 7,3 | ND | 16,0 | ND | 12,6 |
| Alimentação | 5,3 | ND | 16,2 | ND | 15,9 |
| Edição e edição integrada à impressão | 4,2 | ND | 6,8 | ND | 6,6 |
| Atividades de televisão, rádio, cinema e gravação/edição de som e imagem | ND | ND | 14,4 | ND | 12,9 |
| Telecomunicações | 10,4 | ND | 42,2 | ND | 41,7 |
| Desenvolvimento de sistemas e outros serviços de informação | 5,0 | ND | ND | 7,6 | 7,5 |
| Intermediação financeira, seguros e previdência complementar | 9,1 | 9,1 | 21,3 | ND | 20,6 |
| Atividades imobiliárias | 0,8 | ND | 0,8 | ND | 0,8 |
| Atividades jurídicas, contábeis, consultoria e sedes de empresas | 6,7 | ND | 12,4 | ND | 9,5 |
| Serviços de arquitetura, engenharia, testes/análises técnicas e P & D | 4,4 | ND | ND | 4,7 | 4,7 |
| Outras atividades profissionais, científicas e técnicas | 8,9 | ND | 11,2 | ND | 9,7 |
| Aluguéis não-imobiliários e gestão de ativos de propriedade intelectual | 9,9 | ND | 13,4 | ND | 11,8 |
| Outras atividades administrativas e serviços complementares | 6,4 | ND | 11,5 | ND | 7,6 |
| Atividades de vigilância, segurança e investigação | ND | ND | ND | ND | ND |
| Administração pública, defesa e seguridade social | ND | 2,3 | ND | ND | 2,3 |
| Educação pública | ND | 1,3 | ND | ND | 1,3 |
| Educação privada | ND | ND | 4,0 | ND | 4,0 |
| Saúde pública | ND | 2,3 | ND | ND | 2,3 |
| Saúde privada | ND | 5,8 | 5,1 | ND | 5,1 |
| Atividades artísticas, criativas e de espetáculos | 2,6 | ND | 29,8 | ND | 21,9 |
| Organizações associativas e outros serviços pessoais | ND | ND | 5,0 | ND | 4,7 |
| Serviços domésticos | ND | ND | 0,0 | ND | 0,0 |
| **Total** | 7,3 | 2,3 | 19,1 | 11,4 | 13,1 |

Nota: A alíquota tributária efetiva somente é mostrada se a despesa é maior que R$1 bilhão. As alíquotas tributárias efetivas são calculadas dividindo-se a incidência tributária final (tabela 3) pelo valor da transação deduzido do valor da incidência.

**4. Comentários Finais**

Este estudo estimou a incidência final dos tributos sobre mercadorias e transações no Brasil. Os resultados apresentados evidenciam a grande disparidade existente no tratamento tributário de diferentes atividades e produtos. Em particular, destaca-se a forte concentração da arrecadação tributária sobre produtos largamente utilizados como insumos na produção de outros bens e serviços, potencializando distorções no sistema de preços. Observou-se também que a tributação indireta acaba onerando o investimento e as exportações, apesar das isenções legais.

Ressalte-se que princípios básicos de tributação – eficiência econômica, simplicidade e transparência – recomendam que a tributação de bens e serviços se dê por meio de um imposto abrangente sobre o consumo, desonerando as transações entre empresas e o investimento. Isso significa que apenas o consumo das famílias deve ser tributado. É interessante observar que, para gerar a mesma receita total dos tributos considerados neste estudo, um imposto que incida somente sobre o consumo das famílias deve ter uma alíquota (efetiva) de cerca de 25%.

Ao mesmo tempo que as estimativas das alíquotas efetivas aqui apresentas – ao sintetizarem os efeitos de um sistema tributário extremamente complexo e lançar alguma luz sobre a magnitude das distorções resultantes – reforçam nossa convicção na necessidade urgente de uma reforma tributária, este estudo também evidencia o gigantesco desafio que representa imprimir racionalidade ao atual sistema brasileiro de tributação indireta.